\documentclass[preprint,showpacs,preprintnumbers,amsmath,amssymb]{revtex4}
\usepackage{epsfig}
\usepackage{graphicx}
\usepackage{dcolumn}
\usepackage{bm}

\def\beq{\begin{equation}}
\def\eeq#1{\label{#1}\end{equation}}
\def\eeqn{\end{equation}}
\newcommand\iden{\leavevmode\hbox{\small1\normalsize\kern-.33em1}}

\newcommand{\qslash}{\SLASH{q}{.08}}

\newcommand{\SLASH}[2]{\makebox[#2ex][l]{$#1$}/}
\newcommand{\pslash}{\SLASH{p}{.2}}

\newcommand{\bea} {\begin{eqnarray}}
\newcommand{\eea} {\end{eqnarray}}

\newcommand{\Lg}{{\mathcal L}}
\newcommand{\rd}{\partial}

\let\jnfont=\rm
\def\NPB#1,{{\jnfont Nucl.\ Phys.\ B }{\bf #1},}
\def\PLB#1,{{\jnfont Phys.\ Lett.\ B }{\bf #1},}
\def\EPJC#1,{{\jnfont Eur.\ Phys.\ Jour.\ C }{\bf #1},}
\def\PRD#1,{{\jnfont Phys.\ Rev.\ D }{\bf #1},}
\def\PRL#1,{{\jnfont Phys.\ Rev.\ Lett.\ }{\bf #1},}
\def\MPLA#1,{{\jnfont Mod.\ Phys.\ Lett.\ A }{\bf #1},}
\def\JPG#1,{{\jnfont J.\ Phys.\ G }{\bf #1},}
\def\ZPC#1,{{\jnfont Z.\ Phys.\ C }{\bf #1},}
\def\CTP#1,{{\jnfont Commun.\ Theor.\ Phys.\ }{\bf #1},}
\def\JHEP#1,{{\jnfont JHEP \ }{\bf #1},}
\def\NPPS#1,{{\jnfont Nucl.\ Phys.\ Proc.\ Suppl.\ }{\bf #1},}
\def\CPC#1,{{\jnfont Computl.\ Phys.\ Commun.\ }{\bf #1},}
\def\APPB#1,{{\jnfont Acta\ Phys.\ Polon.\ B }{\bf #1},}
\def\JKPS#1,{{\jnfont J.\ Korean\ Phys.\ Soc. }{\bf #1},}

\def\EPL#1,{{\jnfont Europhys.\ Lett. }{\bf #1},}

\begin{document}

\title{\ \\[10mm] The lepton flavor violating
decays $Z\to l_i l_j$ \\ in the simplest little Higgs model}

\author{Xiaofang Han \footnotetext{*)
Email address: xfhan@itp.ac.cn}, Lei Wang}

\affiliation{Department of Physics, Yantai University, Yantai
264005, China}

\begin{abstract}
In the simplest little Higgs model the new flavor-changing
interactions between heavy neutrinos and the Standard Model leptons
can generate contributions to some lepton flavor violating decays of
$Z$-boson at one-loop level, such as $Z \to \tau^{\pm}\mu^{\mp}$,
$Z\to \tau^{\pm}e^{\mp}$, and $Z \to \mu^{\pm}e^{\mp}$. We examine
the decay modes, and find that the branching ratios can reach
$10^{-7}$ for the three decays, which should be accessible at the
Giga$Z$ option of the ILC.

\end{abstract}



\maketitle

\section{Introduction}
Little Higgs theory \cite{LH} has been proposed as an interesting
solution to the hierarchy problem. So far various realizations of
the little Higgs symmetry structure have been proposed
\cite{otherlh,lst,sst,lht}, which can be categorized generally into
two classes \cite{smoking}. One class use the product group,
represented by the littlest Higgs model \cite{lst}, in which the SM
$SU(2)_L$ gauge group is from the diagonal breaking of two (or more)
gauge groups. The other class use the simple group, represented by
the simplest little Higgs model (SLHM) \cite{sst}, in which a single
larger gauge group is broken down to the SM $SU(2)_L$. The flavor
sector of little Higgs models based on product groups, notably the
littlest Higgs model with T-parity (LHT) \cite{lht}, has been
extensively studied \cite{lht-fla}. Recently, some attentions have
been paid to the flavor sector of SLHM
\cite{slh-fla1,slh-fla2,slh-fla3}.

The lepton flavor violating (LFV) decays of $Z$-boson can be a
sensitive probe for new physics because they are extremely
suppressed in the SM but can be greatly enhanced in new physics
models \cite{lep,lep-lht,quark}. The experimental limits obtained at
LEP \cite{lep-z} are
\begin{eqnarray}
BR(Z \rightarrow \tau^{\pm}\mu^{\mp})&<&1.2\times 10^{-5},\nonumber\\
BR(Z \rightarrow \tau^{\pm}e^{\mp})&<&9.8\times10^{-6},\\
BR(Z \rightarrow \mu^{\pm}e^{\mp})&<&1.7\times 10^{-6}.\nonumber
\end{eqnarray}

The next generation $Z$ factory can be realized in the Giga$Z$
option of the International Linear Collider (ILC) \cite{gigaz}.
About $2 \times 10^9$ $Z$ events can be generated in an operational
year of $10^7 s$ of Giga$Z$. Thus the expected sensitivity of
Giga$Z$ to the LFV decays of $Z$-boson could reach \cite{sen-gigaz}
\begin{eqnarray}
BR(Z \rightarrow \tau^{\pm}\mu^{\mp})&\sim& \kappa\times 2.2\times
10^{-8}, \nonumber\\
BR(Z \rightarrow\tau^{\pm}e^{\mp})&\sim&
\kappa\times 6.5\times10^{-8},\\
BR(Z \rightarrow \mu^{\pm}e^{\mp})&\sim& 2.0\times 10^{-9},\nonumber
\end{eqnarray}
with the factor $\kappa$ ranging from 0.2 to 1.0. Therefore, Giga$Z$
can offer an important opportunity to probe the new physics via the
LFV decays of $Z$-boson.

The SLHM predicts the existence of heavy neutrinos, which have
flavor-changing couplings with the SM leptons mediated respectively
by the SM gauge boson $W^{\pm}$ and the new heavy gauge boson
$X^{\pm}$. These couplings can give great contributions to $Z$-boson
decays $Z \to \tau^{\pm}\mu^{\mp}$, $Z\to \tau^{\pm}e^{\mp}$, and $Z
\to \mu^{\pm}e^{\mp}$ at one-loop level. In this paper, we will
calculate the branching ratios of these decay modes, and compare the
results with the sensitivity of Giga$Z$ and the present experimental
bounds, respectively.

This work is organized as follows. In Sec. II we recapitulate the
SLHM. In Sec. III we study respectively the decays $Z \to
\tau^{\pm}\mu^{\mp}$, $Z\to \tau^{\pm}e^{\mp}$ and $Z \to
\mu^{\pm}e^{\mp}$. Finally, we give our conclusion in Sec. IV.

\section{Simplest little Higgs model}
The SLHM is based on $[SU(3) \times U(1)_X]^2$ global symmetry
\cite{sst}. The gauge symmetry $SU(3) \times U(1)_X$ is broken down
to the SM electroweak gauge group by two copies of scalar fields
$\Phi_1$ and $\Phi_2$, which are triplets under the $SU(3)$ with
aligned VEVs $f_1$ and $f_2$. The uneaten five pseudo-Goldstone
bosons can be parameterized as

\beq
\Phi_{1}= e^{ i\; t_\beta \Theta } \left(\begin{array}{c} 0 \\
0 \\ f_1
\end{array}\right)\;,\;\;\;\;
\Phi_{2}= e^{- \; \frac{i}{t_\beta} \Theta} \left(\begin{array}{c} 0 \\  0 \\
f_2
\end{array}\right)\;,
\label{paramet}
\end{equation}
where
\begin{equation}
   \Theta = \frac{1}{f} \left[
        \left( \begin{array}{cc}
        \begin{array}{cc} 0 & 0 \\ 0 & 0 \end{array}
            & H \\
        H^{\dagger} & 0 \end{array} \right)
        + \frac{\eta}{\sqrt{2}}
        \left( \begin{array}{ccr}
        1 & 0 & 0 \\
        0 & 1 & 0 \\
        0 & 0 & 1 \end{array} \right) \right],
\end{equation}
$f=\sqrt{f_1^2+f_2^2}$ and $t_\beta\equiv \tan\beta= f_2 / f_1$.
Under the $SU(2)_L$ SM gauge group, $\eta$ is a real scalar, while
$H$ transforms as a doublet and can be identified as the SM Higgs
doublet. The kinetic term in the non-linear sigma model is \beq
\label{eq:Lg:gauge0} \Lg_\Phi = \sum_{j=1,2}\left| \left(\rd_\mu + i
g A^a_\mu T^a - i \frac{g_x}{3} B^x_\mu \right) \Phi_j \right|^2,
\end{equation}
where $g_x =g t_W / \sqrt{1-t^2_W /3}$, and $t_W=\tan \theta_W$ with
$\theta_W$ being the electroweak mixing angle. As $\Phi_1$ and
$\Phi_2$ develop their VEVs, the new heavy gauge bosons $Z'$, $Y^0$,
$Y^{0\dagger}$ and $X^{\pm}$ get their masses after eating five
Goldstone bosons, \bea M_X & = &
\frac{gf}{\sqrt{2}}\left(1-\frac{v^2}{4f^2}\right),\nonumber\\
 M_{Z'}&=&\frac{\sqrt{2}gf}{\sqrt{3-t_W^2}}
 \left(1-\frac{3-t_W^2}{c_W^2}\frac{v^2}{16f^2}\right),\nonumber\\
M_{Y}&=&\frac{gf}{\sqrt{2}}. \eea

The gauged $SU(3)$ symmetry promotes the SM fermion doublets into
$SU(3)$ triplets. For each generation of lepton, a heavy neutrino is
added, whose mass is \beq m_{N_i}=fs_\beta \lambda_N^i.
\end{equation}
Where $i=1,2,3$ is the generation index and $\lambda_N^i$ is the
Yukawa coupling constant.

After the EWSB the light and the heavy neutrino of the same family
have the mixing, which is parameterized by
$\delta_v=-\frac{v}{\sqrt{2}ft_\beta}$. The mixing angel $\delta_v$
is experimentally constrained to be small \cite{deltav}, and taken
as a typical upper limit $\delta_v<0.05$ following the ref.
\cite{slh-fla1}. Besides, there is family mixing as long as the
Yukawa matrix of heavy neutrinos and that of leptons are not
aligned. This can induce the lepton flavor-changing interactions of
charged currents proportional to
$V^{ij}_\ell\bar{N}_{Li}\gamma^{\mu}X^{+\mu}\ell_{Lj}$ and $\delta_v
V^{ij}_\ell\bar{N}_{Li}\gamma^{\mu}W^{+\mu}\ell_{Lj}$, where
$V^{ij}_\ell$ is the mixing matrix \cite{smoking,slh-fla1,slh-fla2}.

\section{The LFV decays $Z \to
\tau^{\pm}\mu^{\mp}$, $Z\to \tau^{\pm}e^{\mp}$ and $Z \to
\mu^{\pm}e^{\mp}$}

In the SLHM, the Feynman diagrams for $Z \to \mu^{\pm}e^{\mp}$ can
be depicted by the Fig. 1, and the diagrams for $Z \to
\tau^{\pm}\mu^{\mp}$, $Z\to \tau^{\pm}e^{\mp}$ are same as Fig.1,
but replacing $\mu$ and $e$ with the corresponding final particles.
For the 't Hooft-Feynman gauge, the flavor-changing interactions
between the heavy neutrino and lepton, mediated by the gauge bosons
(Goldstone bosons) $X^{\pm}$ ($x^{\pm}$) and $W^{\pm}$
($\phi^{\pm}$), can contribute to these decays. The relevant Feynman
rules can be found in \cite{slh-fla1}.

The calculations of the loop diagrams in Fig. 1 are straightforward.
Each loop diagram is composed of some scalar loop functions
\cite{Hooft} which are calculated by using LoopTools \cite{Hahn}.
The analytic expressions from our calculation are presented in
Appendix A.

\begin{figure}[tb]
\begin{center}
 \epsfig{file=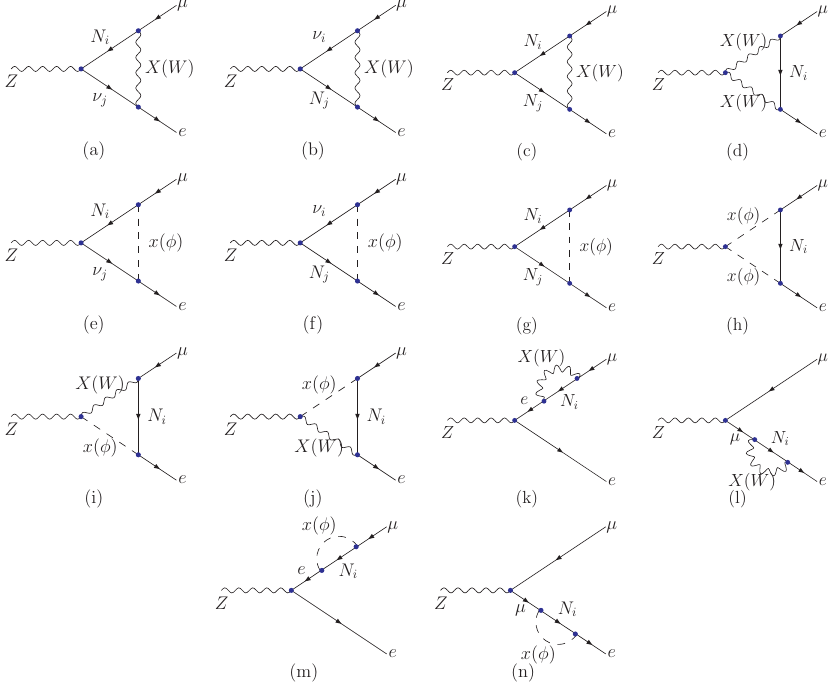,width=15cm}
\end{center}
\caption{Feynman diagrams for $Z\to \mu^{+}e^-$ in the SLHM.}
\label{zffy}
\end{figure}

The SM input parameters relevant in our study are taken as ref.
\cite{pdg}. The free SLHM parameters involved are $f, ~t_\beta$,
 the heavy neutrino mass $m_{N_i}$ $(i=1,2,3)$, and the mixing matrix $V_\ell$
which can be parameterized with standard form. To simply our
calculations, we take the parameters \cite{vl}\beq
s_{12}=\sqrt{0.3},~~s_{13}=\sqrt{0.03},~~s_{23}=\frac{1}{\sqrt{2}},~~\delta_{13}=65^{\circ},
\end{equation}
which is consistent with the experimental constraints on the PMNS
matrix \cite{vl-exp}, and $\delta_{13}$ is taken to be equal to the
CKM phase. To satisfy the present experimental bounds of $Br(\mu\to
e\gamma)$ and $Br(\mu\to eee)$, the mass splitting of the first and
the second heavy neutrinos must be very small \cite{slh-fla1}. So in
this paper we will take $m_{N_1}=m_{N_2}=m_1=400$ GeV and
$m_{N_3}=m_3$ in the range of 500 GeV-3000 GeV. Ref. \cite{sst}
shows that the LEP-II data requires $f
> 2$ TeV. In our numerical calculation we will take several values
of $t_\beta$ for $f = 2$ TeV, $f = 4$
TeV and $f = 5.6$ TeV, respectively.

\begin{figure}[tb]
\begin{center}
 \epsfig{file=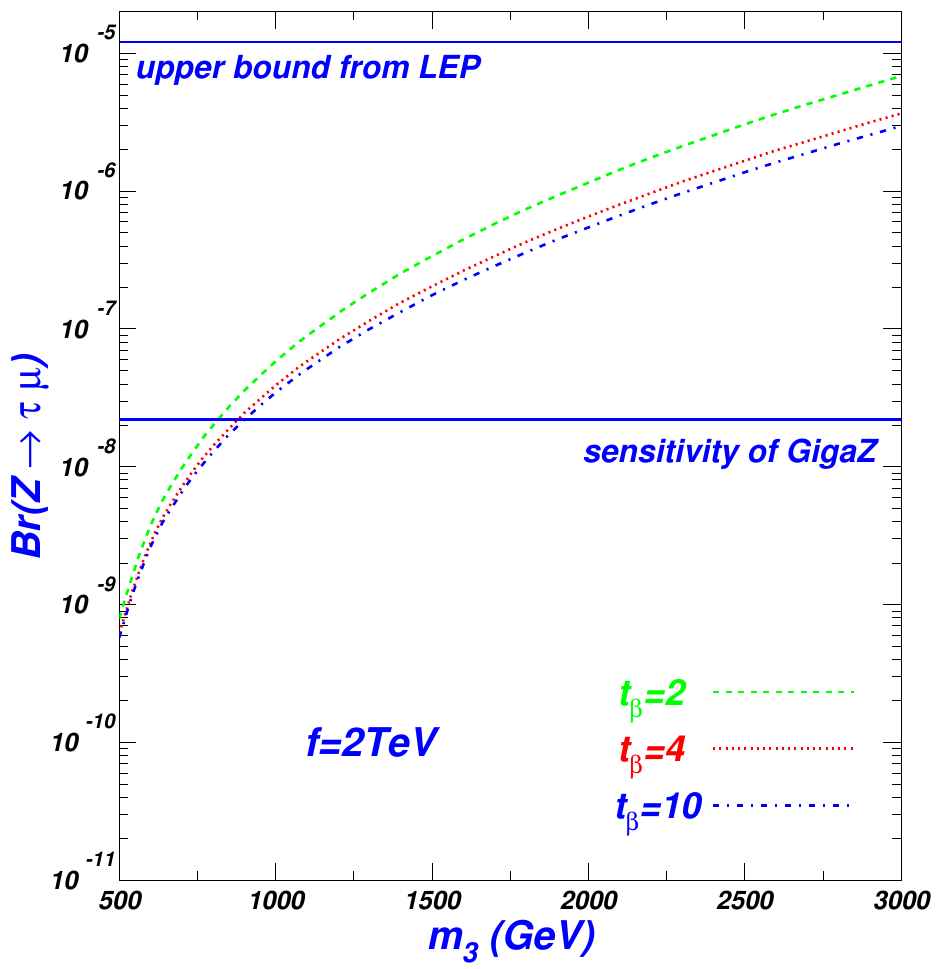,width=5.55cm}
 \epsfig{file=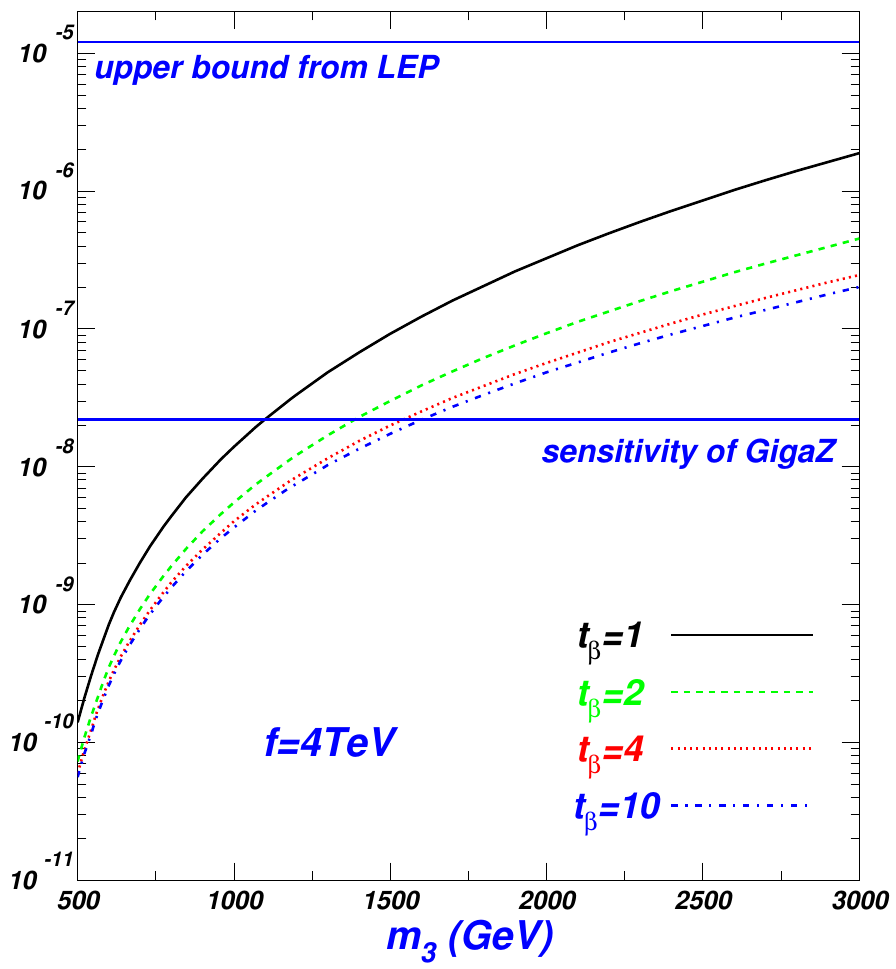,width=5.3cm}
  \epsfig{file=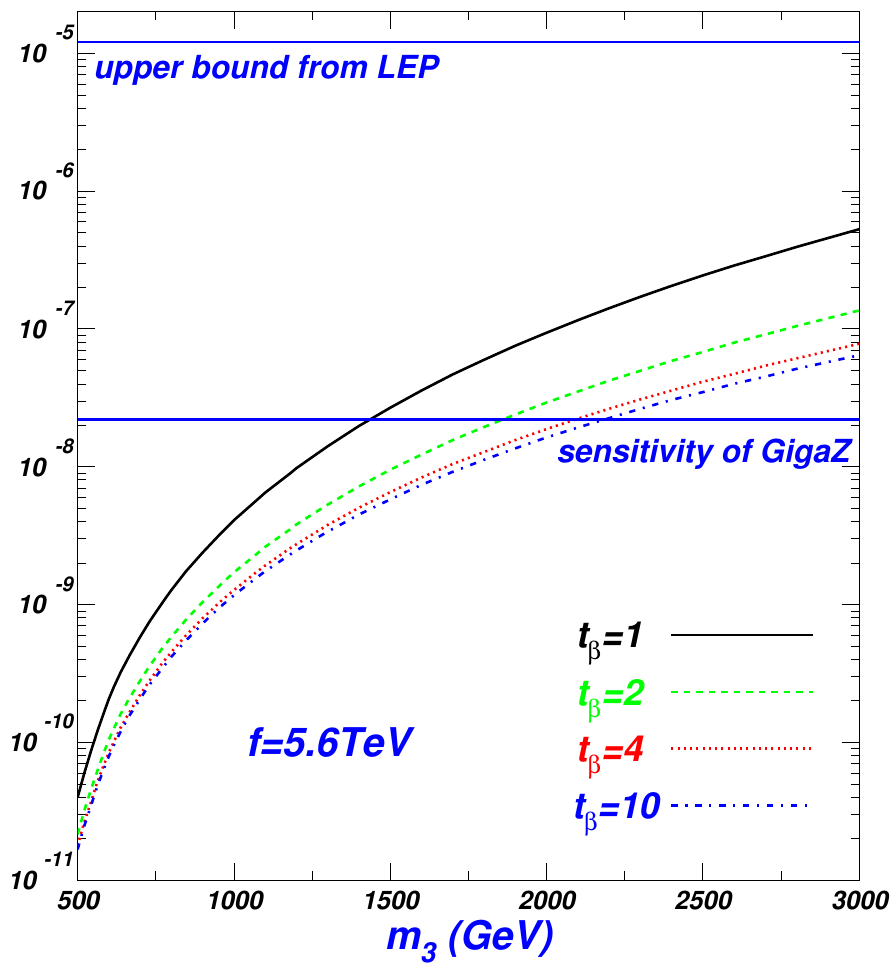,width=5.3cm}
\end{center}
\vspace{-1.0cm} \caption{The branching ratios of $Z \to
\tau^{\pm}\mu^{\mp}$ versus $m_3$.} \label{tu}
\end{figure}
\begin{figure}[tb]
\begin{center}
 \epsfig{file=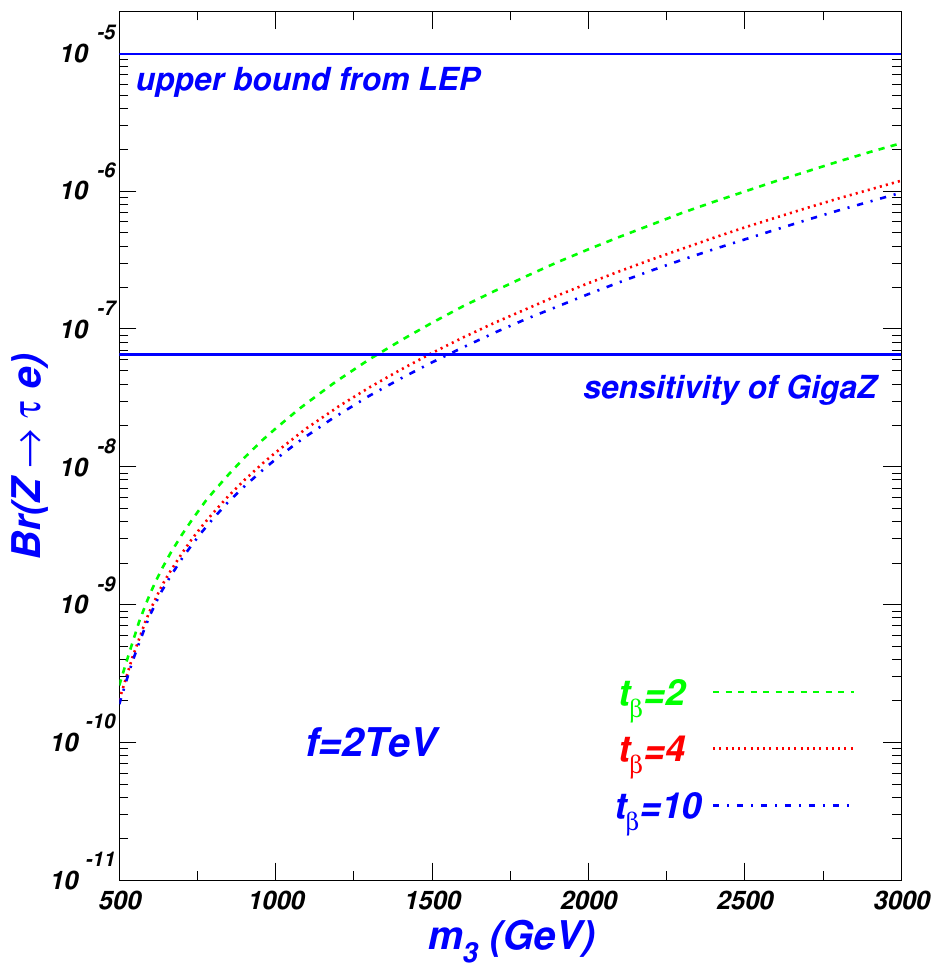,width=5.55cm}
 \epsfig{file=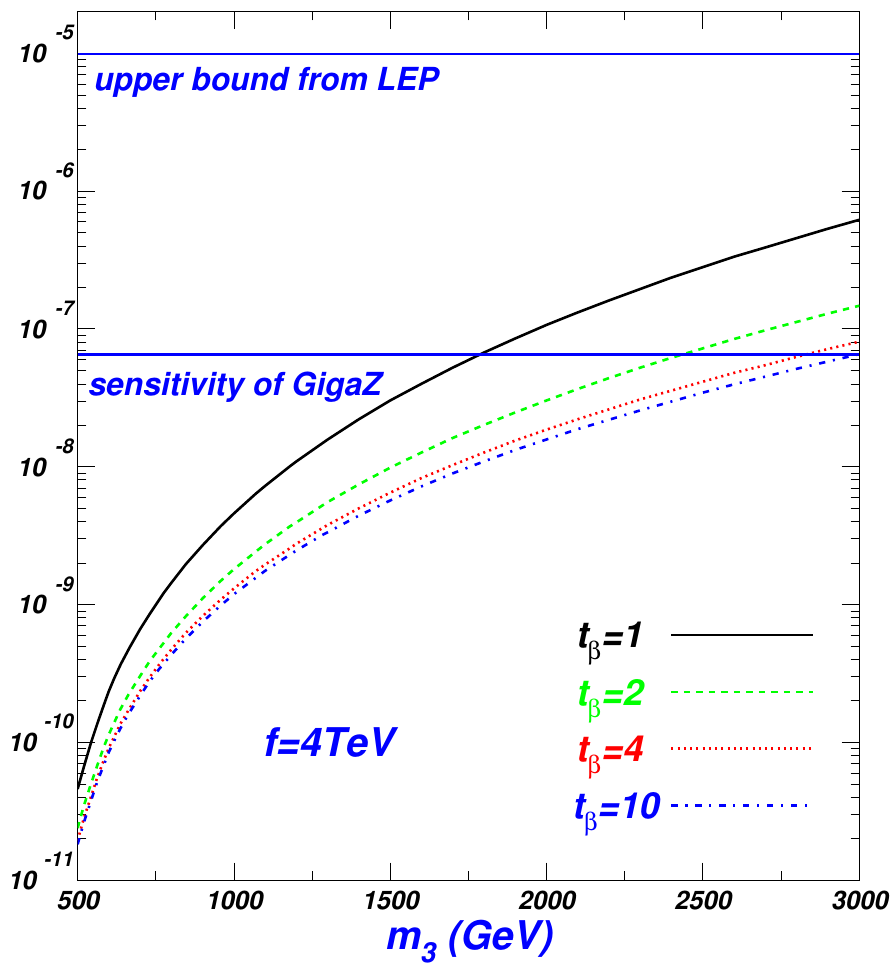,width=5.3cm}
  \epsfig{file=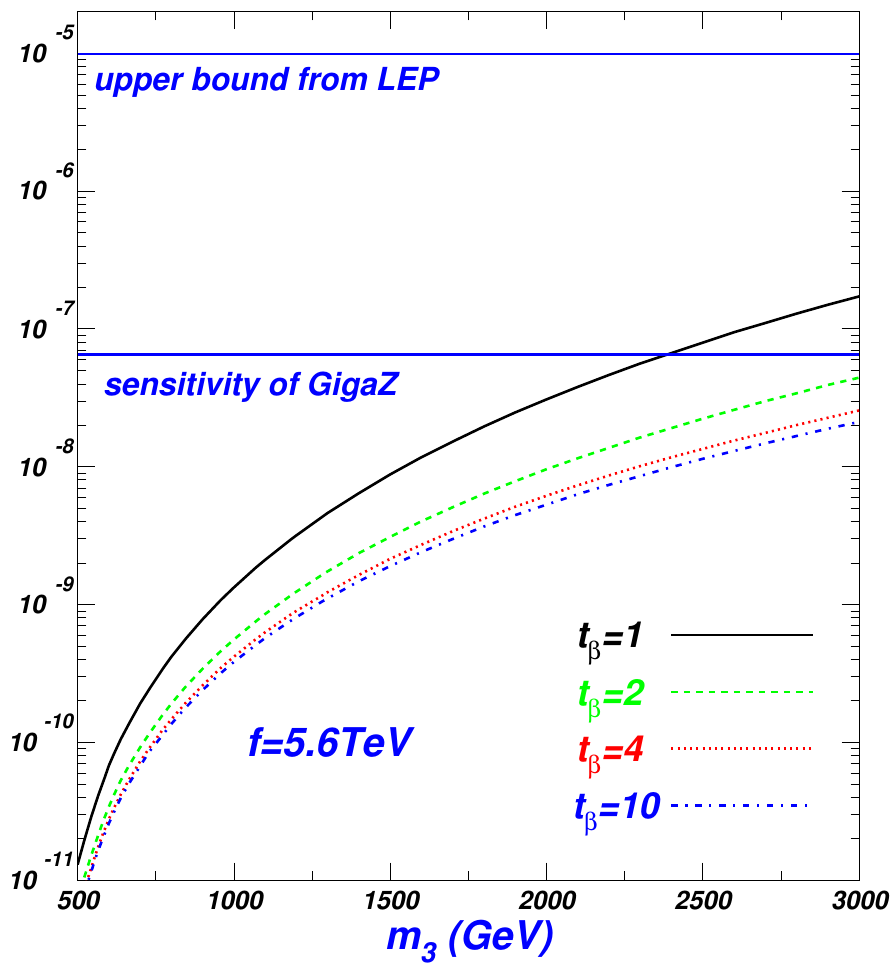,width=5.3cm}
\end{center}
\vspace{-1.0cm} \caption{The branching ratios of $Z\to
\tau^{\pm}e^{\mp}$ versus $m_3$.} \label{te}
\end{figure}
\begin{figure}[tb]
\begin{center}
 \epsfig{file=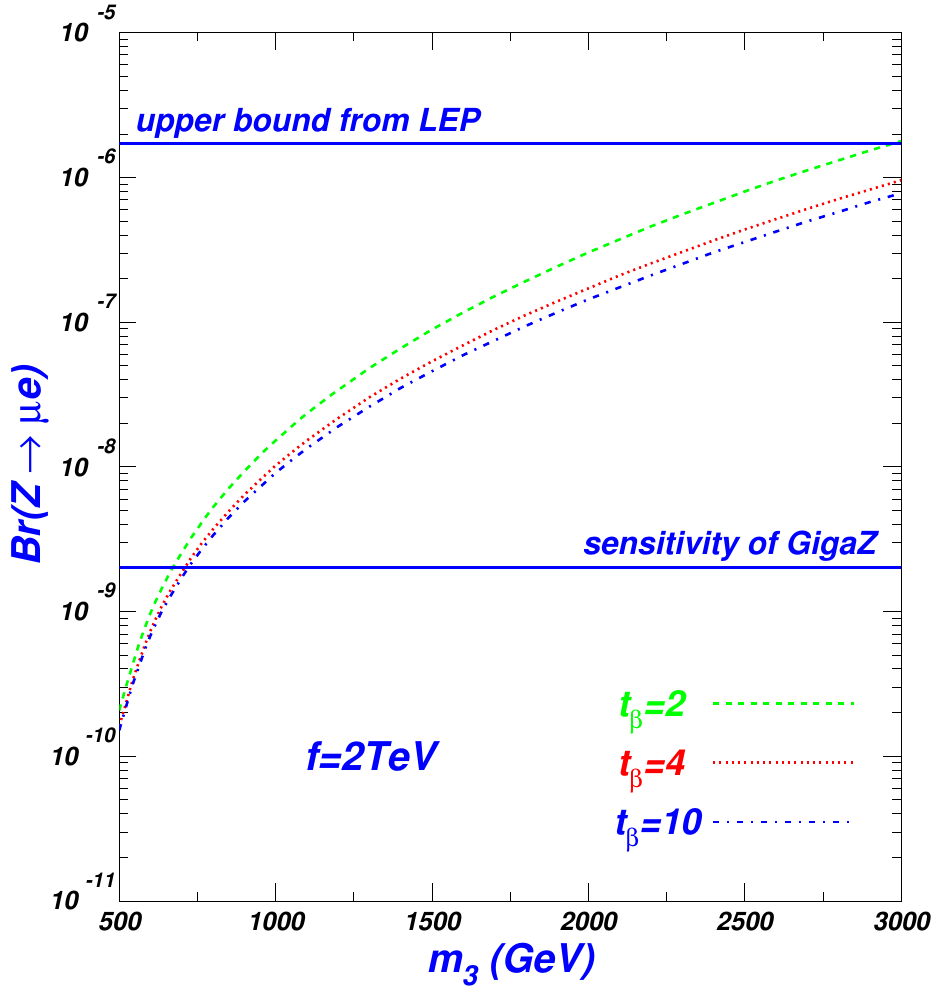,width=5.55cm}
 \epsfig{file=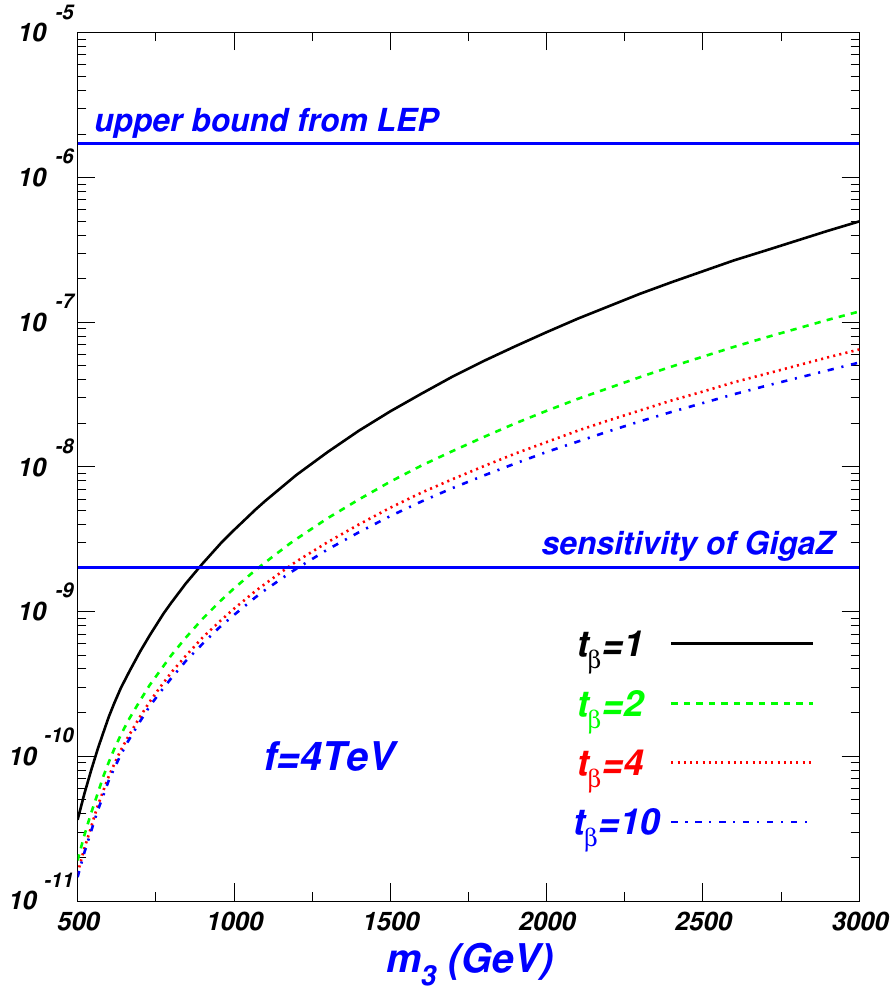,width=5.3cm}
  \epsfig{file=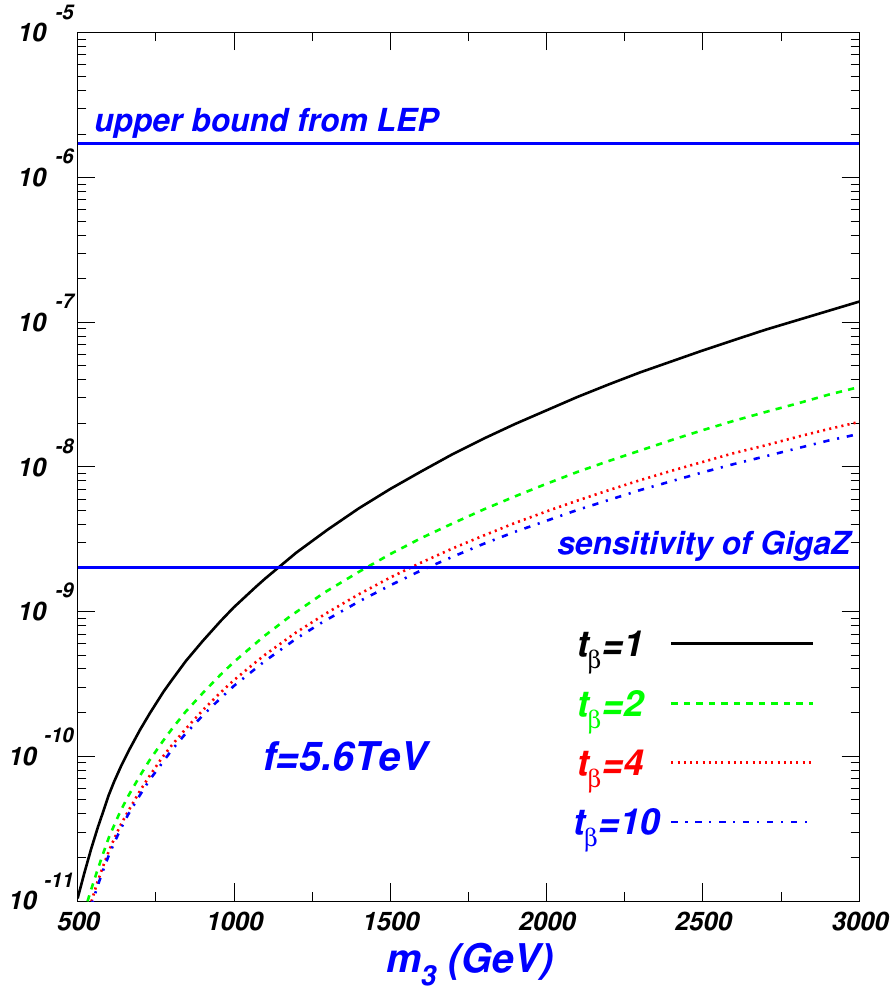,width=5.3cm}
\end{center}
\vspace{-1.0cm} \caption{The branching ratios of $Z \to
\mu^{\pm}e^{\mp}$ versus $m_3$.} \label{ue}
\end{figure}

In Fig. \ref{tu}, Fig. \ref{te} and Fig. \ref{ue}, we plot the decay
branching ratios of $Z \to \tau^{\pm}\mu^{\mp}$, $Z\to
\tau^{\pm}e^{\mp}$ and $Z \to \mu^{\pm}e^{\mp}$ versus $m_3$ for
$f=2$ TeV, $f=4$ TeV and $f=5.6$ TeV, respectively. We find that the
branching ratios increase with the mass of the third generation
heavy neutrino. The reason is that the decays are enhanced by the
large mass splitting $m_3 - m_1$, which increases as $m_3$ gets
large since we have fixed the value of $m_1$. Besides, the branching
ratios drop as the scale $f$ or $t_\beta$ get large, and the reason
is that the lepton flavor-changing couplings
$\bar{N}_{Li}\gamma^{\mu}W^{+\mu}\ell_{Lj}$ and
$\bar{N}_{i}\phi^{+}\ell_{j}$ are proportional to
$\delta_v=-\frac{v}{\sqrt{2}ft_\beta}$.

Fig. \ref{tu}, Fig. \ref{te} and Fig. \ref{ue} show the branching
ratios of $Z \to \tau^{\pm}\mu^{\mp}$, $Z\to \tau^{\pm}e^{\mp}$ and
$Z \to \mu^{\pm}e^{\mp}$ are below the present experimental upper
bounds, respectively. However, the ratios can be enhanced to reach
the sensitivity of the Giga$Z$. For $f=2$ TeV, $t_\beta=4$ and
$m_3=2$ TeV, the branching ratios can reach $10^{-7}$ for $Z \to
\tau^{\pm}\mu^{\mp}$, $Z\to \tau^{\pm}e^{\mp}$ and $Z \to
\mu^{\pm}e^{\mp}$, which exceed much the sensitivity of Giga$Z$. In
the LHT, all the three ratios can reach $10^{-6}$ \cite{lep-lht}.
Therefore, the LFV decays of $Z$-boson may be accessible at Giga$Z$,
and thus may serve as a probe of the little Higgs models.

\section{Conclusion}
In the framework of the simplest little Higgs model, we studied the
LFV decays $Z \to \tau^{\pm}\mu^{\mp}$, $Z\to \tau^{\pm}e^{\mp}$ and
$Z \to \mu^{\pm}e^{\mp}$. In the parameter space allowed by current
experiments, the branching ratios of the three decays can exceed
respectively much the sensitivity of Giga$Z$, which should be
accessible at the Giga$Z$ option of the ILC. Therefore, the
measurement of these rare decays at the Giga$Z$ may serve as a probe
of the simplest little Higgs model.

\section*{Acknowledgment}
This work was supported in part by the National Natural Science
Foundation of China (NNSFC) under grant No. 11005089, and by the
Foundation of Yantai University under Grant Nos. WL10B24 and
WL09B31.

\appendix

\section{The effective coupling of $Z\mu^+ e^-$}
Here we take the effective coupling of $Z\mu^+ e^-$ for example. The
other two couplings $Z\tau^+ \mu^-$ and $Z\tau^+ e^-$ can be
obtained via some corresponding replacement of the analytic
expressions for $Z\mu^+ e^-$. The effective coupling of $Z\mu^+ e^-$
is given by
\begin{eqnarray}
\Gamma_{Z\mu e}^{\alpha}&=&
\Gamma_{VF_{1}F_{2}}^{\alpha}\big[X(W),N_i,\nu_j\big]
+\Gamma_{VF_{1}F_{2}}^{\alpha}\big[X(W),\nu_i,N_j\big]
+\Gamma_{VF_{1}F_{2}}^{\alpha}\big[X(W),N_i,N_j\big]\nonumber\\
&&+\Gamma_{FVV}^{\alpha}\big[N_i,X(W),X(W)\big]
+\Gamma_{SF_{1}F_{2}}^{\alpha}\big[x(\phi),N_i,
\nu_j\big] +\Gamma_{SF_{1}F_{2}}^{\alpha}\big[x(\phi),\nu_i, N_j\big]\nonumber\\
&&+\Gamma_{SF_{1}F_{2}}^{\alpha}\big[x(\phi),N_i, N_j\big]
+\Gamma_{FSS}^{\alpha}\big[N_i,x(\phi),x(\phi)\big]
+\Gamma_{FVS}^{\alpha}\big[N_i,X(W),x(\phi)\big]\nonumber\\
&&+\Gamma_{FSV}^{\alpha}\big[N_i,x(\phi),X(W)\big]
+\Gamma_{self(k)}^{\alpha}\big[X(W),N_i\big]
+\Gamma_{self(l)}^{\alpha}\big[X(W), N_i\big]\nonumber\\
&&+\Gamma_{self(m)}^{\alpha}\big[x(\phi),N_i\big]
+\Gamma_{self(n)}^{\alpha}\big[x(\phi),N_i\big],
\end{eqnarray}
where the particles in the square brackets represent the particles
which contribute to the vertex, and $\Gamma_{self(k-n)}^{\alpha}$
correspond to the vertexes in Fig. 1$(k-n)$. The self-energy and
vertex contributions in the above equation are given by \small
\begin{eqnarray}
\Gamma_{VF_{1}F_{2}}^{\alpha} &=&\frac{i}{16\pi^{2}}
\big[(d_{1}Z_{R}^{f}P_{L}+c_{1}Z_{L}^{f}P_{R})
(-2C_{\sigma\rho}\gamma^{\sigma}\gamma^{\alpha}\gamma^{\rho}-2\gamma^{\alpha})
(c_{2}P_{L}+d_{2}P_{R}) \nonumber\\ & &
-2(\qslash_{e}+\qslash_{\mu})\gamma^{\alpha}C_{\beta}\gamma^{\beta}
(c_{1}c_{2}Z_{L}^{f}P_{L}+d_{1}d_{2}Z_{R}^{f}P_{R})
+4m_{F_{2}}(c_{2}d_{1}Z_{R}^{f}P_{L}\nonumber\\
& &+c_{1}d_{2}Z_{L}^{f}P_{R})C_{\alpha}
+4m_{F_{1}}(c_{2}d_{1}Z_{L}^{f}P_{L}+c_{1}d_{2}Z_{R}^{f}P_{R})C_{\alpha}
+2m_{F_{1}}C_{0}(d_{1}Z_{L}^{f}P_{L}\nonumber\\ & &
+c_{1}Z_{R}^{f}P_{R})
(2(q_{e}+q_{\mu})^{\alpha}-m_{F_{2}}\gamma^{\alpha})(c_{2}P_{L}+d_{2}P_{R})\big]
(q_e,q_{\mu},m_{F1},m_V,m_{F2}),\\
\Gamma_{FVV}^{\alpha}
&=&\frac{ig_{VVV}}{16\pi^{2}}(d_{1}P_{L}+c_{1}P_{R})\big\{
-4C_{\alpha\beta}\gamma^{\beta}+\gamma^{\alpha}-2C_{\beta}\gamma^{\beta}(q_{e}+q_{\mu})^{\alpha}
+2(4m_{F}-2\qslash_{\mu})C_{\alpha} \nonumber\\ & &
+(4m_{F}-2\qslash_{\mu})(q_{e}+q_{\mu})^{\alpha}C_{0}
-\big[C_{\sigma\rho}g^{\sigma\rho}-\frac{1}{2}\big]\gamma^{\alpha}-C_{\beta}\gamma^{\beta}
(\qslash_{\mu}+m_{F})\gamma^{\alpha} -\pslash_{Z}C_{\beta}\gamma^{\beta}\gamma^{\alpha}\nonumber\\
& &-\pslash_{Z}(\qslash_{\mu}+m_{F})\gamma^{\alpha}C_{0}
-\big[C_{\sigma\rho}g^{\sigma\rho}-\frac{1}{2}\big]\gamma^{\alpha}
+C_{\beta}\gamma^{\alpha}\gamma^{\beta}
(\pslash_{Z}-\qslash_{e}-\qslash_{\mu})
-\gamma^{\alpha}(\qslash_{\mu}+m_{F})C_{\beta}\gamma^{\beta}\nonumber\\
& &
+\gamma^{\alpha}(\qslash_{\mu}+m_{F})(\pslash_{Z}-\qslash_{e}-\qslash_{\mu})C_{0}\big\}
(c_{2}P_{L}+d_{2}P_{R})
(q_{\mu},q_e,m_V,m_F,m_V),\\
\Gamma_{SF_{1}F_{2}}^{\alpha} &=&\frac{i}{16\pi^{2}}
\big[C_{\sigma\rho}\gamma^{\sigma}\gamma^{\alpha}\gamma^{\rho}
(a_{2}b_{1}Z_{R}^{f}P_{L}+a_{1}b_{2}Z_{L}^{f}P_{R})
+\frac{1}{2}\gamma^{\alpha}(a_{2}b_{1}Z_{R}^{f}P_{L}+a_{1}b_{2}Z_{L}^{f}P_{R})
 \nonumber\\ & &
+C_{\beta}\gamma^{\beta}\gamma^{\alpha}(a_{1}Z_{L}^{f}P_{L}+b_{1}Z_{R}^{f}P_{R})
(\qslash_{e}+\qslash_{\mu}+m_{F_{2}})(a_{2}P_{L}+b_{2}P_{R})
\nonumber\\
& & +m_{F_{1}}\gamma^{\alpha}
(b_{1}b_{2}Z_{L}^{f}P_{L}+a_{1}a_{2}Z_{R}^{f}P_{R})C_{\beta}\gamma^{\beta}
+m_{F_{1}}\gamma^{\alpha}(b_{1}Z_{L}^{f}P_{L}\nonumber\\
& & +a_{1}Z_{R}^{f}P_{R})
(\qslash_{e}+\qslash_{\mu}+m_{F_{2}})(a_{2}P_{L}+b_{2}P_{R})C_{0}\big]
(q_e,q_{\mu},m_{F1},m_S,m_{F2}),
\end{eqnarray}
\begin{eqnarray}
\Gamma_{FSS}^{\alpha} &=&-\frac{ig_{VSS}}{16\pi^{2}}\big\{
-2C_{\alpha\beta}\gamma^{\beta}(a_{2}b_{1}P_{L}+a_{1}b_{2}P_{R})
-(q_{e}+q_{\mu})^{\alpha}C_{\beta}\gamma^{\beta}(a_{2}b_{1}P_{L}+a_{1}b_{2}P_{R})
\nonumber\\ & &
+\big[-2C_{\alpha}-(q_{e}+q_{\mu})^{\alpha}C_{0}\big]\qslash_{\mu}
(a_{2}b_{1}P_{L}+a_{1}b_{2}P_{R})
+m_{F}\big[-2C_{\alpha}\nonumber\\
& &
-(q_{e}+q_{\mu})^{\alpha}C_{0}\big](a_{1}a_{2}P_{L}+b_{1}b_{2}P_{R})\big\}
(q_{\mu},q_e,m_S,m_F,m_S),\\
\Gamma_{FVS}^{\alpha} &=&-\frac{ig_{VVS}}{16\pi^{2}}
\gamma^{\alpha}(c_{1}P_{L}+d_{1}P_{R})\big[C_{\beta}\gamma^{\beta}
+(\qslash_{\mu}+m_{F})C_{0}\big]\nonumber\\
& & \times (a_{2}P_{L}+b_{2}P_{R})
(q_{\mu},q_e,m_S,m_F,m_V),\\
\Gamma_{FSV}^{\alpha} &=&\frac{ig_{VVS}}{16\pi^{2}}
(a_{1}P_{L}+b_{1}P_{R})\big[C_{\beta} \gamma^{\beta}+(\qslash_{\mu} + m_{F}) C_{0}\big] \gamma^{\alpha} \nonumber\\
& & \times (c_{2}P_{L}+d_{2}P_{R})
(q_{\mu},q_e,m_V,m_F,m_S),\\
\Gamma_{self(k)}^{\alpha} &=&-\frac{ig}{16\pi^{2}c_W
(q_{\mu}^{2}-m_{e}^{2})}
\gamma^{\alpha}\big[(-\frac{1}{2}+s_W^2)P_{L}+s_W^2P_{R}\big](\qslash_{\mu}+m_{e})
\big[(2B_{\beta}\gamma^{\beta}+(2B_{0}\nonumber\\
& & -1)\qslash_{\mu})
(c_{1}c_{2}P_{L}+d_{1}d_{2}P_{R})-2m_{F}(2B_{0}-1)(c_{2}d_{1}P_{L}+c_{1}d_{2}P_{R})\big] (q_{\mu},m_V,m_F),\\
\Gamma_{self(l)}^{\alpha} &=&-\frac{ig}{16\pi^{2}c_W
(p_{e}^{2}-m_{\mu}^{2})}\big[(2B_{\beta}\gamma^{\beta}
+(2B_{0}-1)\pslash_{e})
(c_{1}c_{2}P_{L}+d_{1}d_{2}P_{R})-2m_{F}(2B_{0}\nonumber\\
& & -1)(c_{2}d_{1}P_{L}+c_{1}d_{2}P_{R})\big]
(\pslash_{e}+m_{\mu})\gamma^{\alpha}
\big[(-\frac{1}{2}+s_W^2)P_{L}+s_W^2P_{R}\big]
(p_e,m_V,m_F), \\
\Gamma_{self(m)}^{\alpha} &=&\frac{ig}{16\pi^{2}c_W
(q_{\mu}^{2}-m_{e}^{2})}
\gamma^{\alpha}\big[(-\frac{1}{2}+s_W^2)P_{L}+s_W^2P_{R}\big]
(\qslash_{\mu}+m_{e})
\big [(B_{\beta}\gamma^{\beta} \nonumber\\
& & +\qslash_{\mu} B_{0})(a_{2}b_{1}P_{L}+a_{1}b_{2}P_{R})
+m_{F}B_{0}(a_{1}a_{2}P_{L}+b_{1}b_{2}P_{R})\big] (q_{\mu},m_S,m_F),
\\\Gamma_{self(n)}^{\alpha} &=&\frac{ig}{16\pi^{2}c_W
(p_{e}^{2}-m_{\mu}^{2})}\big[(B_{\beta}\gamma^{\beta}+\pslash_{e}
B_{0})(a_{2}b_{1}P_{L}+a_{1}b_{2}P_{R})
+m_{F}B_{0}(a_{1}a_{2}P_{L} \nonumber\\
& & +b_{1}b_{2}P_{R})\big]
(\pslash_{e}+m_{\mu})\gamma^{\alpha}\big[(-\frac{1}{2}+s_W^2)P_{L}+s_W^2P_{R}\big]
(p_e,m_S,m_F),
\end{eqnarray}
\normalsize where $q_{\mu}=-p_{\mu}$, $q_{e}=-p_{e}$ and
$P_{L,R}=(1\mp\gamma_{5})/2$. The functions $B$ and $C$ are 2- and
3-point Feynman integrals \cite{Hahn}, and their functional
dependence is indicated in the bracket following them. The tensor
loop functions can be expanded as the scalar functions \cite{Hahn}.
In our calculation the contraction of Lorentz indices is performed
numerically. The parameters appearing above are from
\begin{eqnarray*}
V\bar{e}f&:&i\gamma^{\mu}(c_1 P_L+d_1 P_R),\hspace{3cm}
V\bar{f}\mu:i\gamma^{\mu}(c_2 P_L+d_2 P_R),\nonumber\\
S\bar{e}f&:&a_1 P_L+b_1 P_R,\hspace{4cm}
S\bar{f}\mu:a_2 P_L+b_2 P_R,\nonumber\\
ZS^+ S^-&:&ig_{VSS} (p_{S^+}^{\mu}-p_{S^-}^{\mu}),\hspace{2.5cm}
ZV^+ S^-:g_{VVS}g^{\mu\nu},\nonumber\\
Z_{\rho}V^{+}_\mu V^{-}_\nu
&:&-ig_{VVV}[(p_{_{V^+}}-p_{_{V^-}})^\rho g^{\mu\nu}+
(p_Z-p_{_{V^+}})^\nu g^{\mu\rho}+(p_{_{V^-}}-p_Z)^\mu g^{\nu\rho}],\nonumber\\
Z\bar{f_1}f_2&:&i\gamma^{\mu}(Z_L^f P_L+Z_R^f P_R),
\end{eqnarray*}
where $V$ represents gauge bosons and $S$ represents scalar
particles. These couplings represent the seven different classes of
vertices involved in our calculation. In each class of vertices, the
parameters $a_1$, $b_1$, $a_2$, $b_2$, $c_1$, $d_1$, $c_2$, $d_2$,
$g_{VSS}$, $g_{VVS}$, $g_{VVV}$, $Z_L^f$ and $Z_R^f$ take different
values for different concrete coupling. The analytic expressions of
these parameters can be found in \cite{slh-fla1}.

\end{document}